\documentclass[10pt,a4paper,twocolumn]{article}

\input{pisika.dat}

\usepackage{indentfirst}
\usepackage{multirow}
\usepackage{makecell}
\usepackage{float}
\usepackage{stfloats}
\usepackage{booktabs}
\usepackage{comment}
\usepackage{times}
\begin{document}

\providecommand{\ShortAuthorList}[0]{} 
\title{Measurements of decay branching fractions of the Higgs boson to hadronic final states at the CEPC}
\author[1,3,5]{Xiaotian Ma}
\author[1,2]{Zuofei Wu\thanks{Corresponding author: zuofei.wu@cern.ch, co-first author}}
\author[1,4]{Jinfei Wu}
\author[1,5]{Yanping Huang\thanks{Corresponding author: huangyp@ihep.ac.cn}}
\author[1,5]{Gang Li}
\author[1,5]{Manqi Ruan}
\author[2]{F\'{a}bio L. Alves}
\author[2]{Shan Jin}
\author[1,3]{Ligang Shao}
\affil[1]{Institute of High Energy Physics, Chinese Academy of Sciences, Beijing 100049, China}
\affil[2]{School of Physics, Nanjing University, Nanjing, 210093, China}
\affil[3]{School of Physical Sciences, University of Chinese Academy of Sciences, Beijing, China}
\affil[4]{China Center of Advanced Science and Technology, Beijing,  100190, China}
\affil[5]{Center for High Energy Physics, Henan Academy of Sciences, Zhengzhou, China}

\date{\dateline{October 17, 2024}}

\begin{abstract}
\noindent
The Circular Electron Positron Collider (CEPC) is a large-scale particle accelerator designed to collide electrons and positrons at high energies. One of the primary goals of the CEPC is to achieve high-precision measurements of the properties of the Higgs boson, facilitated by the large number of Higgs bosons that can be produced with significantly low contamination. The measurements of Higgs boson branching fractions into $b\overline{b} /c\overline{c} /gg$ and $\tau\overline{\tau} /WW^{*} /ZZ^{*} $, where the $W$ or $Z$ bosons decay hadronically, are presented in the context of the CEPC experiment, assuming a scenario with 5600 fb$^{-1}$ of collision data at a center-of-mass energy of 240 GeV. In this study the Higgs bosons are produced in association with a $Z$ boson, with the $Z$ boson decaying into a pair of muons $(\mu^{+}\mu^{-})$, which have high efficiency and high resolution. In order to separate all decay channels simultaneously with high accuracy, the Particle Flow Network (PFN), a graph-based machine learning model, is considered. The precise classification provided by the PFN is employed in measuring the branching fractions using the migration matrix method, which accurately corrects for detector effects in each decay channel. The statistical uncertainty of the measured branching ratio is estimated to be 0.55\% in $H\to b\overline{b}$ final state, and approximately 1.5\%-16\% in $H\to c\overline{c} /gg/\tau\overline{\tau}/WW^{*} /ZZ^{*} $ final states. In addition, the main sources of systematic uncertainties to the measurement of the branching fractions are discussed.


\DOI{} 
\end{abstract}

\maketitle
\thispagestyle{titlestyle}

\section{Introduction}\label{sec:intro}
The discovery of the Higgs boson by the ATLAS and CMS collaborations at the Large Hadron Collider (LHC)  in July 2012 \cite{Aad_2012, 201230} marked a breakthrough in particle 
physics, providing deeper insights into the Standard Model (SM). While SM has been successful in describing the fundamental building blocks of matter and their interactions, several unanswered questions remain, such as the origin of dark matter and the inability to unify all fundamental forces. As a promising gateway to new physics, precise measurements of the Higgs boson’s properties are essential for testing the Standard Model (SM) and uncovering potential hints of physics beyond the Standard Model (BSM). 
\par
In comparison with the LHC, which relies on high-energy proton-proton ($pp$) collision, a lepton collider offers more energy control and significantly lower pileup contamination (average number of $pp$ interactions per beam crossing), serving as a Higgs factory. Several lepton colliders have been proposed with the aim of reconfirming the discovery of the Higgs-like particle and studying the properties of Higgs boson with high precision, including CLIC \cite{aa8818d58bd64b099b8d539f81fef1fc}, FCC-ee  \cite{e524ba048a324d559571fffa99d38781} and ILC \cite{baer2013international}. Among the aforementioned colliders, the Circular Electron
Positron Collider (CEPC) \cite{thecepcstudygroup2018cepc1, thecepcstudygroup2018cepc2} was proposed by the Chinese High Energy Physics Community in 2012. It is designed to operate at a center-of-mass energy of 240 GeV to 250 GeV with an integrated luminosity of 5600 fb$^{-1}$. The main Higgs production process in CEPC will be via associated production with a $Z$ boson, $e^{+}e^{-}  \to ZH$, where the $Z$ boson is expected to undergo further decay. 
\par
According to theoretical predictions, the branching fractions for the decay of a 125 GeV Higgs boson into $b\overline{b}$, $c\overline{c}$, $gg$, $\tau\overline{\tau}$, $WW^{*}$, $ZZ^{*}$ are 57.7\%, 2.91\%, 8.57\%, 6.32\%, 21.5\% and 2.64\%, respectively \cite{InclusiveObservables,DifferentialDistributions,HiggsProperties}. The Higgs boson decay into $b\overline{b}$, $WW^{*}$, $ZZ^{*}$ were studied by the ATLAS Collaboration using a 13 TeV $pp$ Run 2 dataset collected at a center-of-mass energy of 13 TeV with a luminosity of 139 fb$^{-1}$ at the LHC. The branching fractions were measured to be $0.53\pm0.08$, $0.257^{+0.026}_{-0.024}$, $0.028\pm0.003$, respectively \cite{ATLAS:2022vkf}. 
\par 
The work presented here focuses on the determination of the branching fractions of the Higgs boson decaying into a pair of $b$-quark or $c$-quark, gluons, $\tau\overline{\tau}$, $WW^{*}$ or $ZZ^{*}$ in associated $Z(\mu^+\mu^-)H$ production, where the $W$ or $Z$ bosons decay hadronically, at the CEPC with a center-of-mass energy of 240 GeV and integrated luminosity of 5600 fb$^{-1}$. The branching fraction measurements for $H\to b\overline{b} /c\overline{c} /gg/\tau\overline{\tau}/WW^{*} /ZZ^{*}$ will be conducted simultaneously considering the major background sources. Since the dominant decay modes of $WW^{*}$ and $ZZ^{*}$ are hadronic, all the six processes result primarily in final states with jets, making it challenging to distinguish between them. Such difficulty is addressed by employing the Particle Flow Network (PFN) \cite{Komiske_2019}, which is used for jet tagging, due to its ability to separate these processes. In contrast with traditional jet tagging methods based on QCD theory, which measure branching fractions channel by channel, PFN achieves separation of all channels in a single implementation with high accuracy.

\par
This paper is organized as follows: Section 2 provides a brief description of the collider and the MC simulations. Event selection requirements are detailed in Section 3. Section 4 discusses the modeling using Particle Flow Networks, with their performance evaluated in Section 5. The procedure for determining the branching fractions is explained in Section 6, followed by the results in Section 7, where the measurements and their associated statistical and systematic uncertainties are discussed. A brief summary is given in Section 8.

\section{CEPC detector and simulation samples}
The CEPC is a circular electron positron collider with total circumference of 100 km. The center of mass energy in CEPC could reach the $Z$ pole (91.2 GeV), the $WW$ threshold (161 GeV) and the Higgs factory (240 GeV). The CEPC detector employs a highly granular calorimetry system to separate the particle showers, and a low material tracking system to minimize the interaction of the final state particles in the tracking material. It contains a vertex detector with high spatial resolution, a Time Projection Chamber (TPC), a silicon tracker, a silicon-tungsten sampling Electromagnetic Calorimeter (ECAL), and a steel-Glass Resistive Plate Chambers (GRPC) sampling Hadronic Calorimeter (HCAL). The CEPC detector magnet is an iron-yoke-based solenoid which provides an axial magnetic field of 3 Tesla at the interaction point. The outermost part of the detector is a flux return yoke embedded with a muon detector, which identifies muons inside jets. Further details can be found in Ref. \cite{thecepcstudygroup2018cepc2}.

\par
The signal and background events are both generated using the Monte Carlo (MC) generator Whizard 1.95 \cite{Kilian_2011} and Pythia6 \cite{Sj_strand_2006} for the fragmentation and hadronization. The response of the CEPC detector is simulated using a Delphes-based software suite for fast detector simulation \cite{de_Favereau_2014}, according to the performance of the baseline detector in the CEPC CDR \cite{thecepcstudygroup2018cepc2}. The resolution of impact parameter in the $r\phi$ plane is given as:
\begin{equation}
	\label{Equation 1}
	\sigma_{r\phi}=5\oplus\dfrac{10}{p(\mathrm{GeV})\sin^{3/2}\theta} \mathrm{\mu m}.
\end{equation}
The resolution of particle transverse momenta is given as:
\begin{equation}
	\label{equation2}
	\sigma_{\frac{1}{p_{T}}}=2\times10^{-5}\oplus\dfrac{1\times10^{-3}}{p\sin^{3/2}\theta}\mathrm{GeV}^{-1}.
\end{equation}
The energy resolution of photons is:
\begin{equation}
	\label{equation3}
	\dfrac{\sigma_{E}}{E}=0.01\oplus\dfrac{0.16}{\sqrt{E(\mathrm{GeV})}},
\end{equation}
and that of neutral hadrons is:
\begin{equation}
	\label{equation3}
	\dfrac{\sigma_{E}}{E}=0.03\oplus\dfrac{0.50}{\sqrt{E(\mathrm{GeV})}}
\end{equation}

\par 
In this analysis, Higgs production via $ZH$ process is considered to be the dominant process with $Z$ decaying to a pair of muons and Higgs boson decaying  in pairs of $b\overline{b} /c\overline{c} /gg/\tau\overline{\tau}/WW^{*} /ZZ^{*}$ is the signal process, while inclusive decays of $H\to WW^{*}$ and $H\to ZZ^{*}$ are considered. The backgrounds originate from processes with two-fermion and four-fermion final states. The two-fermion background processes include $l\bar{l}$, $\nu\bar{\nu}$ and $q\bar{q}$, referring to final states with leptons (l), neutrinos ($\nu$) and quarks (q). The four-fermion background includes $(ZZ)_h$, $(ZZ)_l$, $(ZZ)_{sl}$,  $(WW)_h$, $(WW)_l$, $(WW)_{sl}$, $(SZ)_l$, $(SZ)_{sl}$, $(SW)_l$, $(SW)_{sl}$, $(mix)_h$ and $(mix)_{l}$, referring to final states with leptons (l), hadrons (h) and semi-leptons (sl). 
 \autoref{tab:Table 1} presents the cross sections of the signal processes. \autoref{tab:Table 2} provides a summary of the detailed decay modes of the two-fermion and four-fermion backgrounds along with their cross sections. 
\begin{table}[h]
	\centering 
	\caption{Cross sections for the Higgs production via $ZH$ process, where $Z$ boson decays to a muon pair and the Higgs boson decays to  $b\overline{b} /c\overline{c} /gg$ and $\tau\overline{\tau}/WW^{*} /ZZ^{*} $, where the $W$ or $Z$ bosons decay hadronically.}\label{tab:Table 1}
	\resizebox{\linewidth}{!}{
		\begin{tabular}{ccc}
			\toprule
			Process & Higgs decays  & Cross section/fb\\
			\midrule
			\multirow{5}{*}{$ZH$ process} & $H\to b\overline{b}$ & 3.91 \\
			& $H\to c\overline{c}$ & 0.20 \\
			& $H\to gg$ & 0.58 \\
			& $H\to \tau\overline{\tau}$ & 0.42\\
			& $H\to WW^{*}$ & 1.46 \\
			& $H\to ZZ^{*}$ & 0.18 \\
			\bottomrule
	\end{tabular}}
\end{table}
\begin{table}[hbtp] 
	\centering 
	\caption{Detailed decay modes for two-fermion ($l\bar{l}$, $\nu\bar{\nu}$ and $q\bar{q}$) and four-fermion ($(ZZ)_h$, $(ZZ)_l$, $(ZZ)_{sl}$, $(WW)_h$, $(WW)_l$, $(WW)_{sl}$, $(SZ)_l$, $(SZ)_{sl}$, $(SW)_l$, $(SW)_{sl}$, $(mix)_h$ and $(mix)_{l}$) backgrounds and their cross sections.}\label{tab:Table 2}
	\resizebox{\linewidth}{!}{
		\begin{tabular}{cccc}
			\toprule
			Category & Name & Decay modes  & Cross section/fb\\ 
			\midrule
			\multirow{11}{*}{\makecell{Two-fermion \\background}}&
			\multirow{3}{*}{$l\bar{l}$} & $e^{+}e^{-} \to e^{+}e^{-}$& 24770.90\\
			&&$e^{+}e^{-} \to \mu^{+}\mu^{-}$&
			5332.71\\
			&&$e^{+}e^{-}\to\tau^{+}\tau^{-}$&
			4752.89\\
			\cline{2-4}
			&\multirow{3}{*}{$\nu\bar{\nu}$} & $e^{+}e^{-} \to \nu_{e}\bar{\nu}_{e}$& 45390.79\\
			&&$e^{+}e^{-} \to \nu_{\mu}\bar{\nu}_{\mu}$&
			4416.30\\
			&&$e^{+}e^{-}\to \nu_{\tau}\bar{\nu}_{\tau}$&
			4410.26\\
			\cline{2-4}
			&\multirow{5}{*}{$q\bar{q}$} & $e^{+}e^{-} \to u\bar{u}$& 10899.33\\
			&&$e^{+}e^{-} \to d\bar{d}$&
			10711.01\\
			&&$e^{+}e^{-}\to c\bar{c}$&
			10862.86\\
			&&$e^{+}e^{-} \to s\bar{s}$& 10737.84\\
			&&$e^{+}e^{-} \to b\bar{b}$&
			10769.78\\
			\midrule
			\multirow{41}{*}{\makecell{Four-fermion \\background}}&
			\multirow{4}{*}{$(ZZ)_h$} & $Z \to c\bar{c},Z \to d\bar{d}/b\bar{b}$& 98.97\\
			&&$ZZ \to d\bar{d}d\bar{d}$&
			233.46\\
			&&$ZZ \to u\bar{u}u\bar{u}$&
			85.68\\
			&&$Z \to u\bar{u},Z \to s\bar{s}/b\bar{b}$ & 98.56\\
			\cline{2-4}
			&\multirow{5}{*}{$(ZZ)_l$} & $Z\to\mu^{+}\mu^{-},Z\to\mu^{+}\mu^{-}$& 15.56\\
			&&$Z\to\tau^{+}\tau^{-},Z\to\tau^{+}\tau^{-}$&
			4.61\\
			&&$Z\to\mu^{+}\mu^{-},Z\to\nu_{\mu}\bar{\nu}_{\mu}$&
			19.38\\
			&&$Z\to\tau^{+}\tau^{-},Z\to\mu^{+}\mu^{-}$&
			18.65\\
			&&$Z\to\tau^{+}\tau^{-},Z\to\nu_{\tau}\bar{\nu}_{\tau}$&
			9.61\\
			\cline{2-4}
			&\multirow{6}{*}{$(ZZ)_{sl}$} & $Z\to\mu^{+}\mu^{-},Z\to d\bar{d}$& 136.14\\
			&&$Z\to\mu^{+}\mu^{-},Z\to u\bar{u}$&
			87.39\\
			&&$Z\to\nu\bar{\nu},Z\to d\bar{d}$&
			139.71\\
			&&$Z\to\nu\bar{\nu},Z\to u\bar{u}$& 84.38\\
			&&$Z\to\tau^{+}\tau^{-},Z\to d\bar{d}$&
			67.31\\
			&&$Z\to\tau^{+}\tau^{-},Z\to u\bar{u}$&
			41.56\\
			\cline{2-4}
			&\multirow{5}{*}{$(WW)_{h}$}&
			$WW\to uubd$& 0.05\\
			&&$WW\to ccbs$&	5.89\\
			&&$WW\to ccds$& 170.18\\
			&&$WW\to cusd$& 3478.89\\
			&&$WW\to uusd$& 170.45\\		
			\cline{2-4}
			&$(WW)_{l}$&
			$WW\to 4leptons$& 403.66\\
			\cline{2-4}
			&\multirow{2}{*}{$(WW)_{sl}$}&
			$W\to\mu\bar{\nu}_{\mu},W\to q\bar{q}$& 2423.43\\
			&&$W\to\tau\bar{\nu}_{\tau},W\to q\bar{q}$& 2423.56\\	
			\cline{2-4}
			&\multirow{6}{*}{$(SZ)_{l}$}&
			$e^{+}e^{-},Z\to e^{+}e^{-}$&  78.49\\
			&&$e^{+}e^{-},Z\to \mu^{+}\mu^{-}$&  845.81\\
			&&$e^{+}e^{-},Z\to \nu\nu$&  28.94\\
			&&$e^{+}e^{-},Z\to \tau^{+}\tau^{-}$&  147.28\\
			&&$\nu^{+}\nu^{-},Z\to \mu^{+}\mu^{-}$&   43.42\\
			&&$\nu^{+}\nu^{-},Z\to \tau^{+}\tau^{-}$&  14.57\\		
			\cline{2-4}
			&\multirow{4}{*}{$(SZ)_{sl}$}&
			$e^{+}e^{-},Z\to d\bar{d}$&  125.83\\
			&&$e^{+}e^{-}, Z\to u\bar{u}$&  190.21\\
			&&$\nu^{+}\nu^{-},Z\to d\bar{d}$&  90.03\\
			&&$\nu^{+}\nu^{-},Z\to u\bar{u}$&  55.59\\	
			\cline{2-4}
			&\multirow{2}{*}{$(SW)_{l}$}&
			$e\nu_{e},W\to \mu\nu_{\mu}$&  436.70\\
			&&$e\nu_{e},W\to \tau\nu_{\tau}$&  435.93\\
			\cline{2-4}
			&$(SW)_{sl}$& $e\nu_{e},W\to qq$&  2612.62\\
			\cline{2-4}
			&\multirow{2}{*}{$(mix)_{h}$}&
			$ZZ/WW\to ccss$&  1607.55\\
			&&$ZZ/WW\to uudd$&  1610.32\\
			\cline{2-4}
			&\multirow{3}{*}{$(mix)_{l}$}&
			$ZZ/WW\to \mu\mu\nu_{\mu}\nu_{\mu}$&  221.10\\
			&&$ZZ/WW\to\tau\tau\nu_{\tau}\nu_{\tau}$&  211.18\\
			&&$SZ/SW\to ee\nu_{e}\nu_{e}$&  249.48\\
			\bottomrule
	\end{tabular}}
\end{table}

\section{Event selection}

\par
 The following criteria are applied to select events for further analysis. Each event must contain at least two oppositely charged tracks, reconstructed as a muon pair ($\mu^{+}\mu^{-}$). The muon candidates in each event are required to be isolated by satisfying $E_{\text{cone}}^{2} <4E_{\mu } +12.2$GeV \cite{Bai_2020}, where $E_{\text{cone}}$ is the sum of energy within a cone ($\cos\theta _{\text{cone}} > 0.98$) around the muon. In cases where more than two muons are selected, the muon pair with the invariant mass closest to the $Z$ boson mass is chosen as the $Z$ candidate, corresponding to a $Z$-mass window of 75 GeV to 105 GeV. The invariant mass of the recoil system, $M_{\mu\mu}^{\text{recoil}}$, against the $Z$ boson candidate is defined as:
 \begin{equation}
 	M_{\mu\mu}^{\text{recoil}} = \sqrt{(\sqrt{s}-E_{\mu^{+}}-E_{\mu^{-}})^{2}-(\overrightarrow{P_{\mu^{+}}}+\overrightarrow{P_{\mu^{-}}})^{2}}
 \end{equation}
 where $\sqrt{s}=240$ GeV while $E$ and $\overrightarrow{P}$ represent the energy and momentum of the muons, respectively. Based on that, $M_{\mu\mu}^{\text{recoil}}$ must fall within the Higgs mass window of 110 GeV to 150 GeV. To further reduce the two-fermion background, the polar angle of muon pair system is required to be in the range of $|\cos\theta_{\mu^{+}\mu^{-}}|<0.996$.
 
 \begin{figure}[hp]
 	\centering
 	\begin{minipage}{1.0\linewidth}
 		\centering
 		\includegraphics[width=1.0\linewidth]{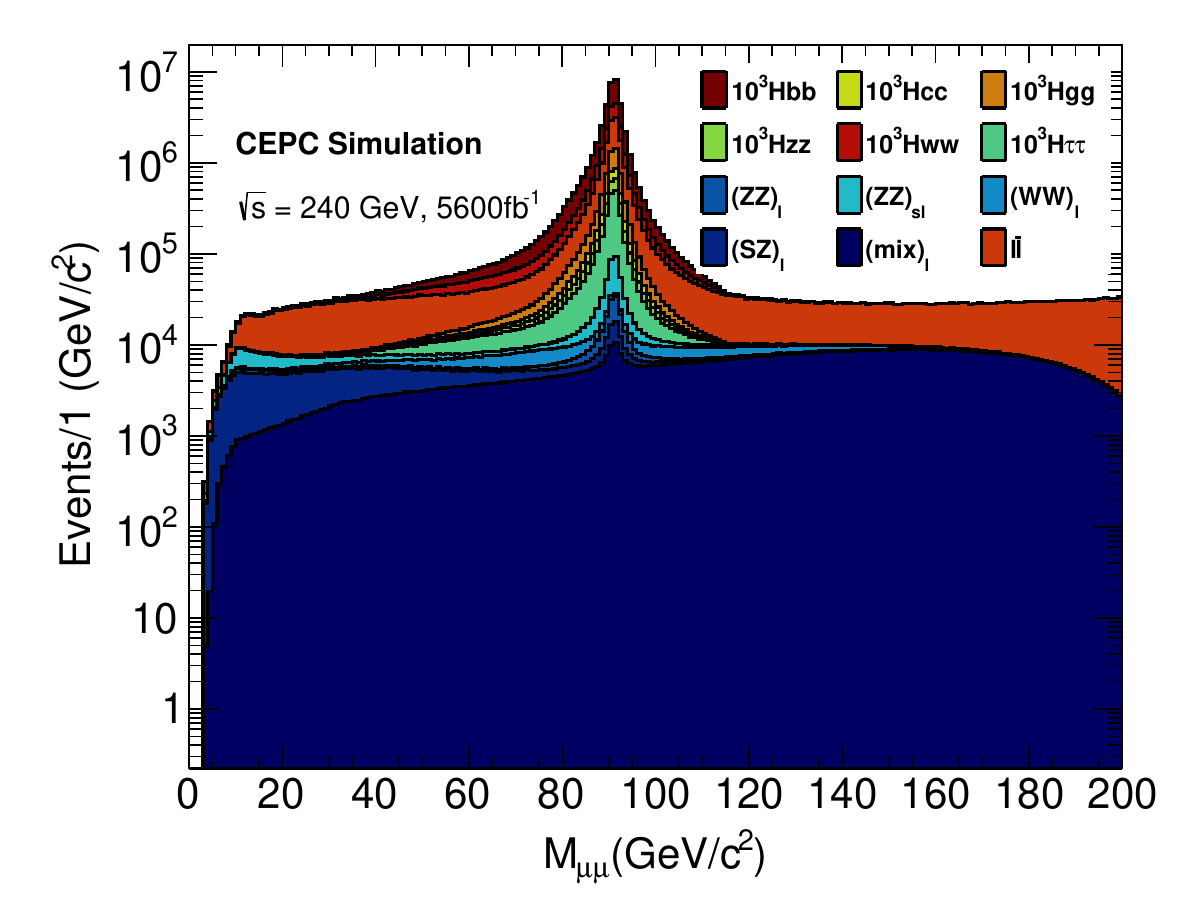}
 		\caption{The invariant mass distributions of the muon pair for signal and background events, after applying the muon pair and isolation selection criteria, are shown. The signal is well preserved, with a high efficiency exceeding 90\%, while background contributions are largely suppressed. Signal events are normalized to 1000 times the expected yields, and background events are normalized to their expected yields in data with an integrated luminosity of 5600 fb$^{-1}$.}
 		\label{fig:Figure 1}
 	\end{minipage}
 \end{figure}
 \begin{figure}[hp]
 	\centering
 	\begin{minipage}{1\linewidth}
 		\centering
 		\includegraphics[width=1.0\linewidth]{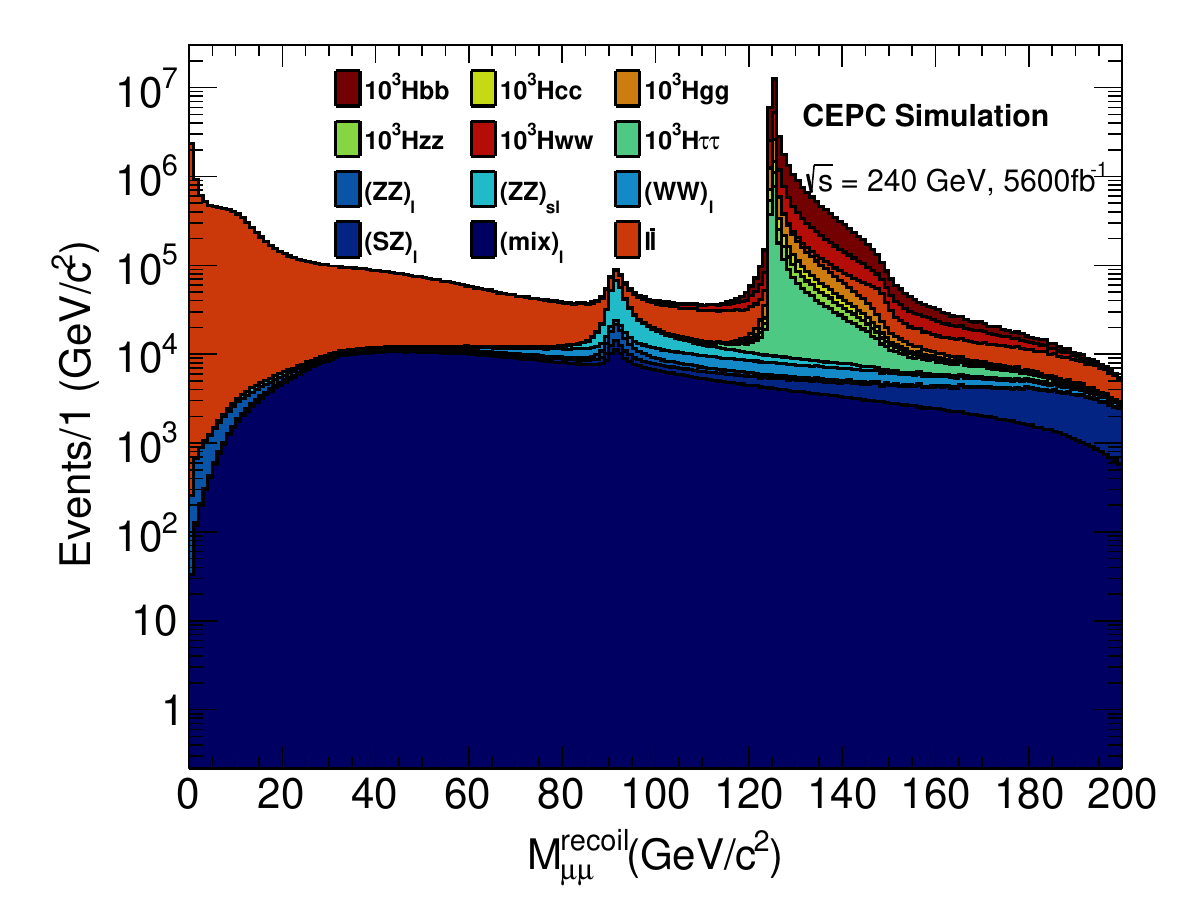}
 		\caption{The invariant mass distributions of the muon pair  recoil system for signal and background events, following the muon pair and isolation selection criteria, are shown. The signal is well preserved, with an efficiency exceeding 90\%, while background contributions are significantly suppressed. Signal events are normalized to 1000 times the expected yields, and background events are normalized to their expected yields in data with an integrated luminosity of 5600 fb$^{-1}$.}
 		\label{fig:Figure 2}
 	\end{minipage}
 \end{figure} 

 \par
  \autoref{fig:Figure 1} shows the invariant mass distribution of the selected muon pair, and \autoref{fig:Figure 2} presents the invariant mass distribution of the muon pair recoil system for both signal and background events, after the isolation and muon pair criteria have been applied. In both distributions, a high signal efficiency of more than 90\% is achieved, while the background contributions are significantly suppressed, following the mass window selections.

 \begin{table*}[ht]
 	\centering
 	\caption{The cutflow selection efficiency is shown for signal and background processes. The relative selection efficiency after each requirement applied and the total selection efficiency for each process are listed.} \label{tab:Table 3}
 	\resizebox{\linewidth}{!}{
 		\begin{tabular}{cccccccccc}
 			
 			\toprule
 		 	&$H\to b\overline{b}$ & $H\to c\overline{c}$ &&$H\to gg$&$H\to \tau\overline{\tau}$&&$H\to WW^{*}$&$H\to ZZ^{*}$&\\
 			\midrule    
 			Simulated events&$1.00\times10^{6}$&$1.00\times10^{6}$&&$1.00\times10^{6}$&$3.72\times10^{5}$&
 			&$1.00\times10^{6}$&$1.00\times10^{6}$&\\
 			Muon pair&94.45\% &94.24\% &&94.17\% &94.94\% &&94.91\% &94.43\%&\\
 			Isolation&91.47\%&92.76\%&&93.31\%&94.47\%&&93.77\%&93.99\%&\\
 			$Z$-mass window&96.28\%&96.41\%&&96.41\%&92.95\%&&93.03\%&95.28\%&\\
 			$H$-mass window&99.64\%&99.66\%&&99.65\%&98.98\%&&98.88\%&99.36\%&\\
 			$|\cos\theta_{\mu^{+}\mu^{-}}|<0.996$&99.66\%&99.66\%&&99.66\%&99.64\%&&99.65\%&99.65\%&\\
 			Total efficiency&82.59\%&83.70\%&&84.14\%&81.95\%&&81.58\%&83.72\%&\\
 			\bottomrule
 			&&&&&&&&&\\
 			\toprule
 			&$l\bar{l}$&$\nu\bar{\nu}$&$q\bar{q}$&$(ZZ)_h$&$(ZZ)_l$&$(ZZ)_{sl}$&$(WW)_{h}$&$(WW)_{l}$&$(WW)_{sl}$\\
 			\midrule    
 			Simulated events&$1.20\times10^{8}$&$3.03\times10^{7}$&$3.03\times10^{7}$&$3.00\times10^{6}$&$1.00\times10^{7}$&
 			$2.60\times10^{7}$&$2.50\times10^{7}$&$2.00\times10^{7}$&$3.00\times10^{7}$\\       
 			Muon pair		&11.95\% 	&0	&0.05\%		&0.08\%	&46.21\% &18.91\% &0.00\%&11.03\%&0.16\%\\
 			Isolation		&91.67\%	&0	&0.40\%	&2.60\%	&74.09\% &66.49\% &0	  &96.46\%&3.68\%\\
 			$Z$-mass window	&41.82\%		&0	&0			&0	&67.68\% &71.45\%  &0	  &34.48\% &17.75\%\\
 			$H$-mass window	&6.55\%		&0	&0			&0	&14.52\%  &15.02\%  &0	  &57.50\% &36.76\%\\
 			$|\cos\theta_{\mu^{+}\mu^{-}}|<0.996$	&90.62\%		&0	&0			&0 &98.83\%  &99.56\%  &0	  &98.85\% &99.15\%\\
 			Total efficiency&0.27\%&0.00\%&0.00\%&0.00\%&3.32\%&1.34\%&0.00\%&2.09\%&0.00\%\\
 			\bottomrule
 			&&&&&&&&&\\
 			\toprule
			&$(SZ)_{l}$ & $(SZ)_{sl}$ &&$(SW)_{l}$&$(SW)_{sl}$&&$(mix)_{h}$&$(mix)_{l}$&\\
			\midrule    
			Simulated events&$8.18\times10^{7}$&$3.20\times10^{6}$&&$3.49\times10^{6}$&$1.05\times10^{7}$&
			&$1.29\times10^{7}$&$1.17\times10^{7}$&\\
			Muon pair&9.92\% &0.02\% &&0 &0.00\% &&0.00\% &29.38\%&\\
			Isolation&44.68\%&0&&0&0&&0&60.77\%&\\
			$Z$-mass window&18.46\%&0&&0&0&&0&13.78\%&\\
			$H$-mass window&31.71\%&0&&0&0&&0&35.94\%&\\
			$|\cos\theta_{\mu^{+}\mu^{-}}|<0.996$&90.02\%&0&&0&0&&0&62.79\%&\\
			Total efficiency&0.36\%&0.00\%&&0.00\%&0.00\%&&0.00\%&0.19\%&\\
\bottomrule
 	\end{tabular}}
 \end{table*}

\par

\autoref{tab:Table 3} presents the event selection efficiencies for various signal and background processes, detailing the efficiency at each selection step relative to the previous requirement. In addition, the total efficiency is defined as the ratio of the number of events satisfying all selection criteria to the total number of events expected from the process considered (signal or background). For signal processes, a high efficiency of over 80\% is observed. In contrast, two-fermion background processes, primarily $l\bar{l}$, exhibit a total efficiency of around 0.3\% and other contributions are negligible. Four-fermion backgrounds, such as $(ZZ)_l$, $(ZZ)_{sl}$ and $(WW)_l$, have total efficiencies of 3.3\%, 1.3\% and 2.1\%, respectively, while $(ZZ)_h$, $(WW)_{h}$, $(WW)_{sl}$ are found to be negligible.

\section{Modeling with Particle Flow Networks}
Machine learning algorithms, particularly those with strong momentum in data analysis, improve their performance as they gain more experience through observational data or interactions with their environment. In particle physics, several neural network models, such as Particle Flow Networks (PFN), Particle Net \cite{Qu_2020} and Particle 
Transformer \cite{qu2022particle} have demonstrated excellent performance in tasks like event classification and jet tagging.
\par
Inspired by point clouds and DeepSet theory \cite{zaheer2018deep}, the Ref. \cite{Komiske_2019} introduced Energy Flow Networks (EFN) and then developed Particle Flow Networks which could accommodate inputs of all information at particle level. This end-to-end 
learning approach eliminates the dependency on jet clustering and e/$\gamma$ isolation. In the DeepSet conception, permutation invariance
and equivariance are essential for handling unordered sets of data. The EFN relies on summation, a symmetric operation that ensures invariance across the elements in the set. PFN defines a mapping for encoding events, defined as $F(\sum_{i} \Phi(p_{i}))$, where 
$p$ represents particle features such as rapidity or transverse momentum, and $\Phi(p)$ is a latent space representation of those features. The function F maps the encoded representations to the network's output. The architecture of the PFN model is defined by the number of layers and neurons within both F and $\Phi$.
\par
In configuring the PFN model, after evaluating various configurations, parameters yielding the best performance were chosen. The function $\Phi(p)$ consists of three layers where the number of neurons in each layer is 64, 64, and 50 neurons. In addition, the function F also contains three layers with the number of neurons set to 64, 64, and 40 neurons. The fully connected layer is directly used in both 
$\Phi$ and F. Each layer uses the ReLU activation function \cite{he2015delving} and adam optimizer \cite{kingma2017adam}. The SoftMax activation function is applied to the output
layer.
\par
  Based on the selection criteria discussed in Section 3, the training process involves a twelve-classification task. The signal includes six distinct Higgs decay channels, while the background contains one two-fermion background class ($l\bar{l}$) and five four-fermion classes ($(ZZ)_{l}$, $(ZZ)_{sl}$, $(WW)_{l}$, $(SZ)_{l}$ and $(mix)_{l}$). 
  During the training procedure, 300,000 events for each process are provided to the model whose weights are all equal to 1, with data split into training, validation and test sets in an 8:1:1 ratio. 
  The PFN is an end-to-end neural network designed to directly utilize the information of the particles to perform event classification. The training variables include the energy of the particle, momentum,
  $\phi$ which is the azimuth angle, $cos\theta$ where $\theta$ is the  polar angle, particle identification number (PID), and impact parameters including $D_0$ and $Z_0$, which represent coordinates in cylindrical coordinate system. 

\par
For the remaining hyperparameters in the training, the number of epoch is set to 200, with a batch size of 1000 and a learning rate of 0.001. 
The loss function uses cross-entropy for multi-class classification problems, while the SoftMax function in the final output layer calculates the score for each class of a given event, which can be used for further analysis.

\section{The model performance}
\begin{figure}[H]
	\centering
	\begin{minipage}{1\linewidth}
		\centering
		\includegraphics[width=1\linewidth]{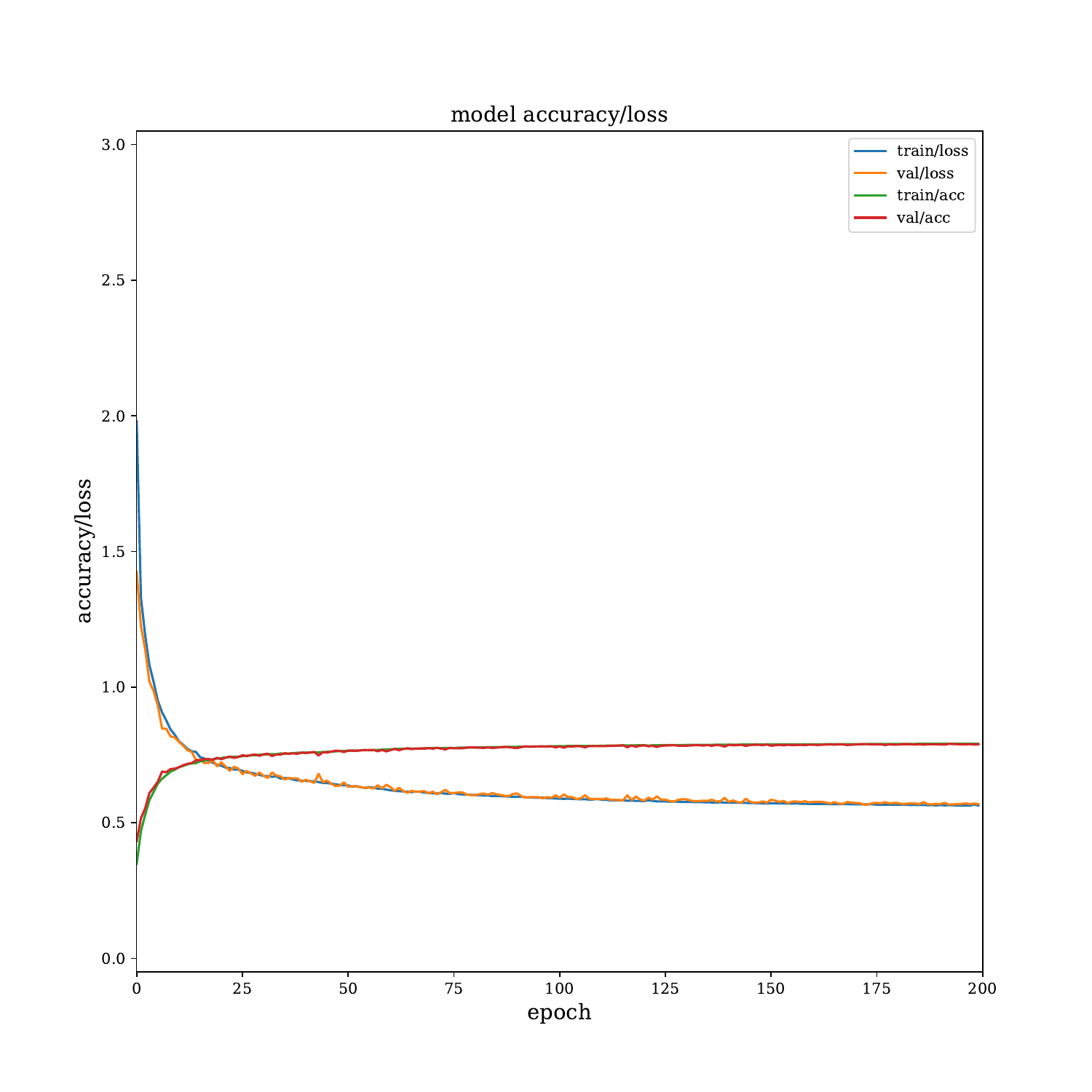}
		\caption{The Loss-accuracy vs epochs curves. The upper two lines are the accuracy curves for the training and validation sets, while the bottom lines are the loss curves for the training and validation set.}
		\label{fig:Figure 3}
	\end{minipage}
\end{figure}

In order to assess the performance of the model, several aspects are considered as described in the following. After each epoch of training, the neural network assesses itself using a validation set, generating a loss-accuracy curve that tracks changes in accuracy throughout the training process. This curve is particularly useful for detecting potential overfitting. 
As shown in \autoref{fig:Figure 3}, the loss and accuracy curves converge towards the end of the training and the high overlap of the training and validation set curves indicates that the model has strong generalization capabilities. 

\par
\begin{figure}[hbt]
	\centering
	\begin{minipage}{1\linewidth}
		\centering
		\includegraphics[width=1\linewidth]{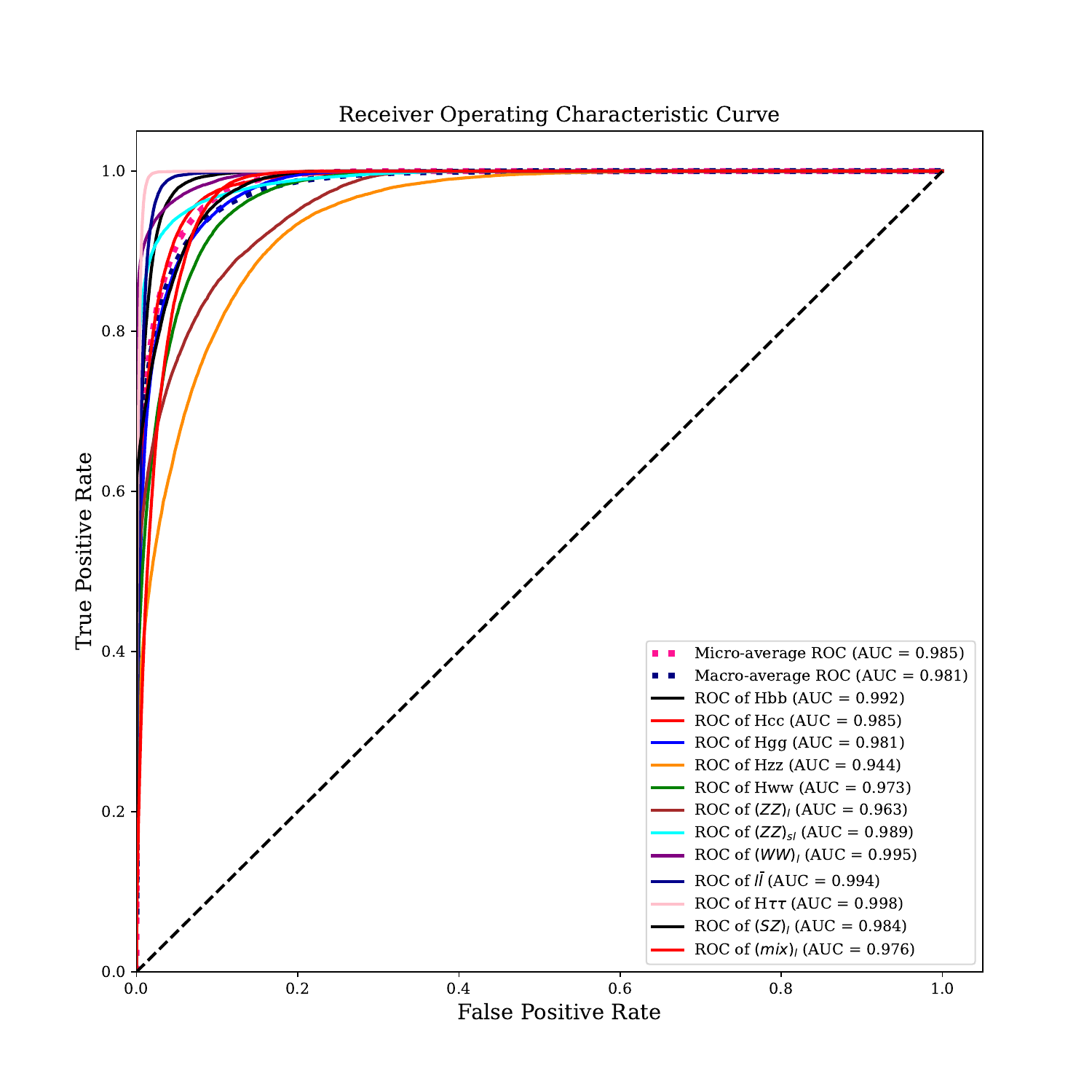}
		\caption{ROC curves for signal and background processes used in classification. The solid lines are the ROC curves of each process considered, and the dashed lines are the ROC curves of the micro and macro average. The dashed black line represents random classification. As can be seen, AUC vaule for each class is above 0.94, indicating a strong classification of the model.}
		\label{fig:Figure 4}
	\end{minipage}
\end{figure}
The Receiver Operating Characteristic Curve (ROC) is a graphical representation of the discriminant power of a classifier model as the threshold is varied. \autoref{fig:Figure 4} depicts the True Positive Rate (TPR) versus the False Positive Rate (FPR) at various discrimination thresholds. The goal of the training is to maximize the TPR while minimizing FPR; therefore the Area Under the Curve (AUC) value serves as an important metric for evaluating the performance of the model. The area under the ROC curve ranges from 0 to 1, where a value of 1 indicates perfect classification and a value of 0.5 suggests a random classification, indicating that the classifier lacks discriminatory power. As can be seen in Figure 4, the AUC vaule for each class is above 0.94, indicating a strong classification performance and the model's ability to effectively distinguish between classes.

\begin{figure*}[hpt]
	\centering
	\begin{minipage}{1\linewidth}
		\centering
		\includegraphics[width=1\linewidth]{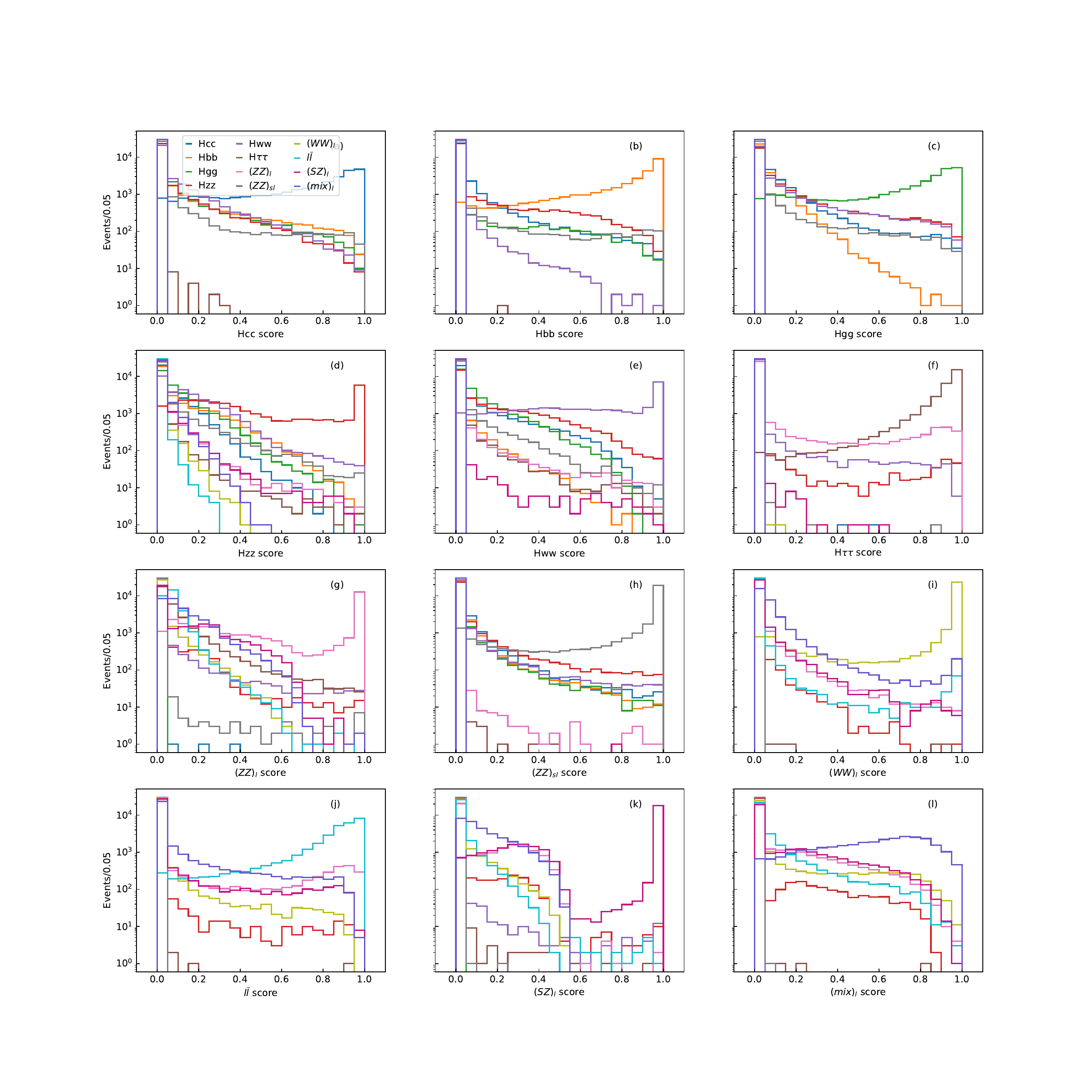}
		\caption{The distributions of classifier outputs for twelve categories are shown. Each histogram represents the probability distribution for processes identified within each category.}
		\label{fig:Figure 5}
	\end{minipage}
\end{figure*}
\par
The classifier outputs are obtained from a nine-unit layer using the SoftMax function. Considering the category $H\to b\overline{b}$ as an example, the SoftMax function computes twelve scores for each event, representing the probability distribution for each process being classified as $H\to b\overline{b}$. 
As illustrated in \autoref{fig:Figure 5} (b), in the region where the score exceeds 0.8, 99\% of the events correspond to the $H\to b\overline{b}$ signal process, while only 1\% of the events originate from the $(ZZ)_{sl}$ background. It can be due to the $Z\to \mu^{+}  \mu ^{-},Z\to u\bar{u}/d\bar{d}$  
processes in the $(ZZ)_{sl}$ background, which have the similar properties with the signal, making the classification more challenging. In addition, the PFN has similar performance in 
other categories. Furthermore, the PFN demonstrates similar performance across other categories. In order to understand the twelve-dimensional more intuitively,  the t-SNE
algorithm \cite{Maaten2008VisualizingDU} is applied to reduce the dimension of the dataset.

\begin{figure}[h]
	\centering
	\begin{minipage}{1\linewidth}
		\centering
		\includegraphics[width=1\linewidth]{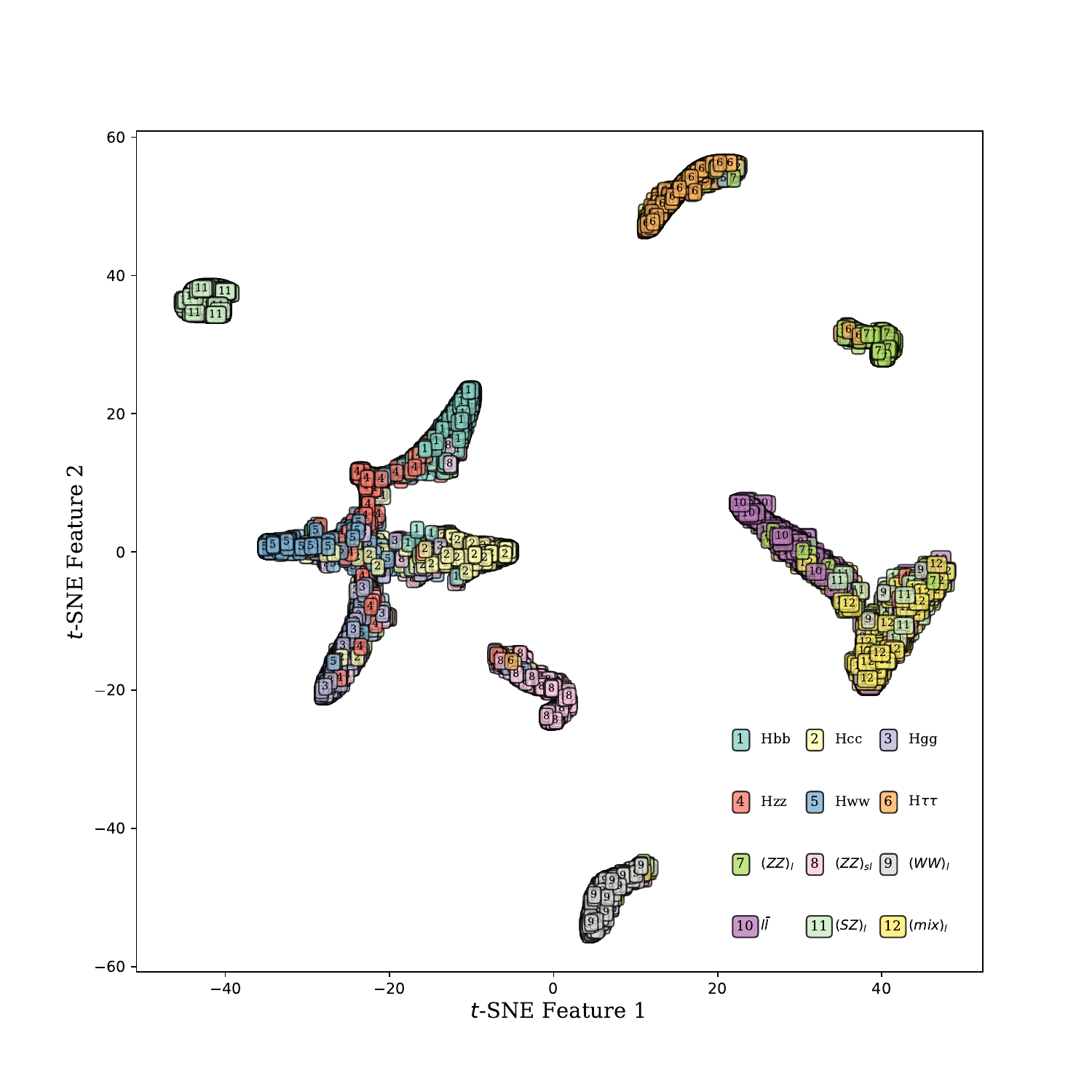}
		\caption{Classification performance visualized using t-SNE algorithm. Different colored squares represent distinct processes, with two t-SNE features corresponding to similarity dimensions. The distance between squares reflects the difference between processes.}
		\label{fig:Figure 6}
	\end{minipage}
\end{figure}

\par
As a non-linear dimension reduction algorithm, t-SNE constructs a similarity matrix and aims to preserve the relationships between data points in both high-dimensional and low-dimensional spaces. The differences in high dimensions are represented as distances in two or three dimensions. As shown in \autoref{fig:Figure 6}, $(WW)_{l}$ and $(SZ)_{l}$ processes are relatively well separated,
while signal process as $H\to c\overline{c}$, $H\to gg$, $H\to WW^{*}$ overlap significantly. In addition, $H\to ZZ^{*}$ process shows similarity to all other signal processes, indicating room for further optimization in model training.

 \par
 \begin{figure}[h]
 	\centering
 	\begin{minipage}{1\linewidth}
 		\centering
 		\includegraphics[width=1\linewidth]{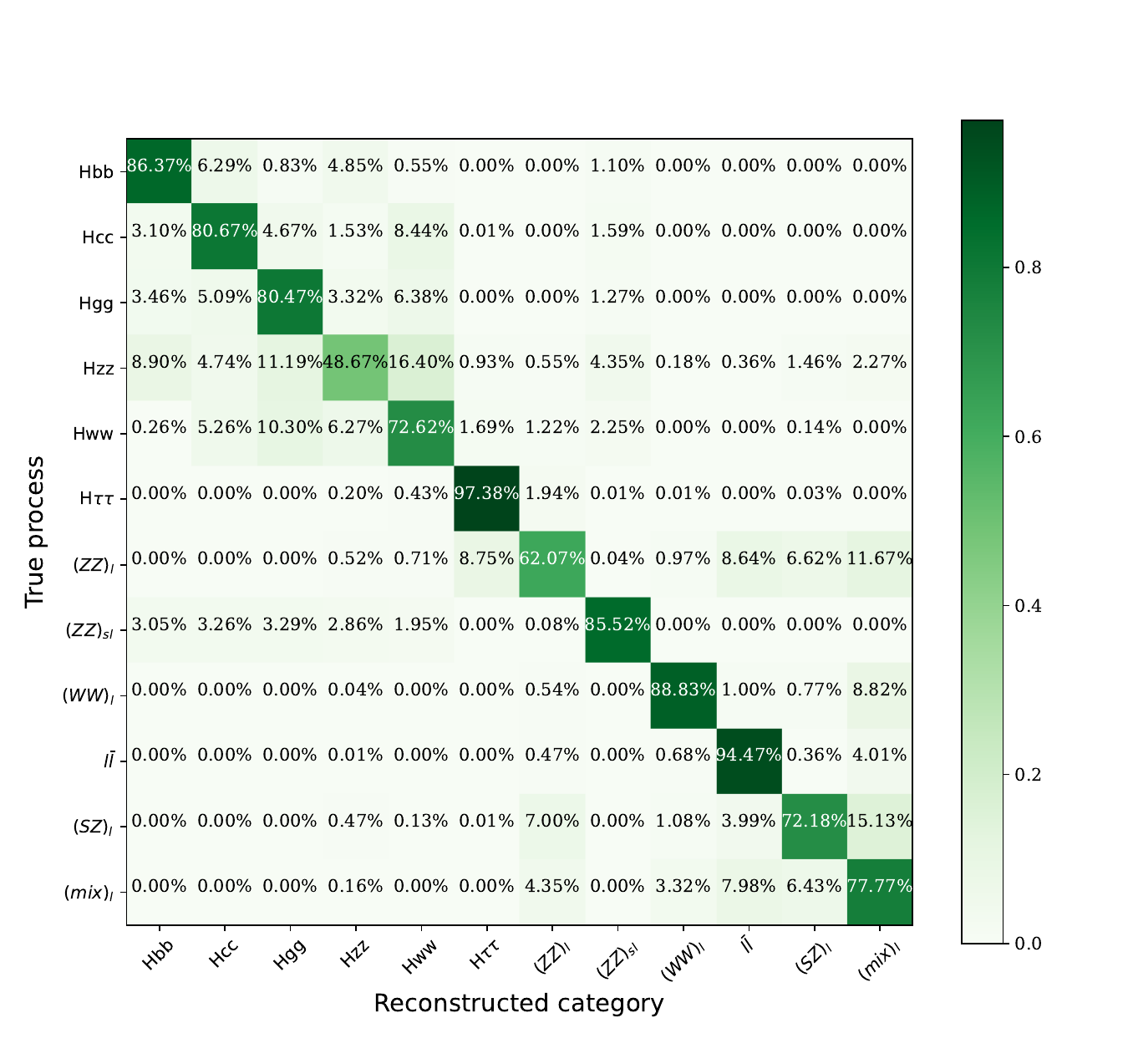}
 		\caption{The migration matrix for the 12 classes is shown. The horizontal axis represents the prediction of the model for each event in the test set, while the vertical axis indicates the true labels. The sum of values in each row equals 1.}
 		\label{fig:Figure 7}
 	\end{minipage}
 \end{figure}
 In supervised learning, the migration matrix is used to compare the classified model’s predictions and the true values. Based on the twelve classification task, there are twelve reconstructed categories, which refer to the process with the highest score for a given event. In \autoref{fig:Figure 7}, the diagonal elements of the matrix represent the correctly classified rates, indicating the purity of each category, while the off-diagonal elements show the misclassification rates. The sum of values in each row equals 1. The decays of $H\to WW^{*}$ and $H\to ZZ^{*}$ are considered inclusively, while the hadronic decays can be well separated from non-hadronic decays by the classifier. The migration matrix reflects the overall high accuracy of the model.

\section{The determination of the branching fractions}
\begin{table*}[hb]
	\centering
	\caption{The measured branching fractions for the Higgs decays along with their statistical uncertainties are shown. The statistical uncertainty ranges from 0.55\% ($H\rightarrow b\bar{b}$) to 15.81\% ($H\rightarrow ZZ^{*}$).} \label{tab:Table 4}
	\resizebox{\linewidth}{!}{
		\begin{tabular}{cccccccccccc}
			
			\toprule
			Higgs boson decay
			&$H\to b\overline{b}$ && $H\to c\overline{c}$ &&$H\to gg$&&$H\to \tau\overline{\tau}$&&$H\to WW^{*}$&&$H\to ZZ^{*}$\\
			\midrule
			branching fraction
			&0.5770&&0.0291&&0.0857&&0.0632&&0.2150&&0.0264\\
			\midrule
			statistical uncertainty &$\pm0.55\%$&&$\pm8.59\%$&&$\pm3.03\%$&&$\pm2.85\%$&&$\pm1.58\%$&&$\pm15.81\%$\\
			\bottomrule
	\end{tabular}}
\end{table*}
The migration matrix contains the information of both correct and incorrect classifications and can be unfolded to represent the generated number of signals \cite{Li_2022}. This matrix method is therefore used to measure the branching fractions of Higgs decays. By considering all signal and background processes, the generated numbers of events for each process can be calculated as in the following: 
\begin{equation}
	\label{eq:law2}
	\begin{bmatrix}
		N_{s1}\\
		N_{s2}\\
		...\\
		N_{b1}\\
		N_{b2}\\
		...
	\end{bmatrix}=\left ( M^{T}_{mig} M_{s} \right  )^{-1} \times \begin{bmatrix}
		n_{s1}\\
		n_{s2}\\
		...\\
		n_{b1}\\
		n_{b2}\\
		...
	\end{bmatrix}
\end{equation}
where $n_{i}$ and $N_{i}$ are the expected and generated number of events of class i, respectively. The $M_{s}$ is a diagonal matrix containing the selection efficiencies, while $M^{T}_{mig}$ denotes the transposed migration matrix: 
\begin{equation}
	\label{eq:law2}
	M^{T}_{mig} =\begin{pmatrix}
		\epsilon_{1,1}  & ... &\epsilon _{12,1} \\
		...&...&...\\
		\epsilon _{1,12}  &... &\epsilon _{12,12} 
	\end{pmatrix}
\end{equation}
where $\epsilon_{ij}$ is the rate at which state i is reconstructed as state j, which is just the corresponding element of the transposed migration matrix.  Besides, $n_{i}$ is obtained from MC samples processed by the PFN model. The branching fraction for each process is then calculated by dividing the corresponding generated number of events by the total number of events from Higgs decays.

\section{Results}
In this analysis, by using the PFN method to separate events in $\mu^{+}\mu^{-}H$ process, the branching fractions of $H\to b\overline{b} /c\overline{c} /gg/\tau\overline{\tau}/WW^{*} /ZZ^{*} $ at the CEPC, with a center-of-mass energy of 240 GeV and luminosity of 5600 fb$^{-1}$, are measured to be 0.5770, 0.0291, 0.0857, 0.0632, 0.2150 and 0.0264, with the statistical uncertainty of $0.55\%$, $8.59\%$, $3.03\%$, $2.85\%$, $1.58\%$ and $15.81\%$, respectively.
\par
The statistical uncertainty is estimated by using toyMC method. The number of events are fluctuated based on a Poisson distribution and then applied to a multinomial distribution according to the migration matrix and selection efficiency. A least squares fit of the measured branching fractions to theoretical fractions is performed 50k times, as shown in \autoref{eq:Equation 6}:
\begin{equation}
	\label{eq:Equation 6}
	\chi^{2}=\sum_{i=1}^{N}\left(\frac{Y_{i}-\eta_{i}}{\sigma_{i}}\right)^{2}
\end{equation} 
where $Y_{i}$ is theoretical branching fraction of process $i$, and $\eta_{i}$ is the measured branching fraction with an error of $\sigma_{i}$. The final results are fitted with guassian function of Higgs decays, where the mean value represents the fitted branching fraction and $\sigma$ denotes the statistic error. The fit results and statistical uncertainties are summarized in \autoref{tab:Table 4}.

\par

\par
To account for the systematic uncertainty, the resolution of transverse momentum of the detector was adjusted by increasing it by 2\% to represent for differences between real data and simulated samples.
By applying the previous PFN model to MC samples generated with updated resolutions, the differences in branching fractions before and after the resolution change are considered as the systematic uncertainty. The systematic uncertainties for the branching fractions are estimated to be 0.21\%, 3.88\%, 2.74\%, 1.39\%, 0.18\% and 19.09\% for $ b\overline{b} /c\overline{c} /gg/\tau\overline{\tau}/WW^{*} /ZZ^{*} $ final states, respectively.

\section{Conclusion}
The Higgs boson branching fractions into $b\overline{b} /c\overline{c} /gg$ and $\tau\overline{\tau}/WW^{*} /ZZ^{*}$, where the $W$ or $Z$ bosons decay hadronically, via the $Z(\mu^{+}\mu^{-})H$ process are studied using the PFN method at a center-of-mass energy of 240 GeV and a luminosity of 5600 fb$^{-1}$ at the CEPC. Simulated samples of "two-fermion" and "four-fermion" processes are considered as backgrounds. The PFN model demonstrates strong performance in classifying different channels and generalizing across processes. The statistical uncertainty of branching fractions of $H\rightarrow b\overline{b} /c\overline{c} /gg/\tau\overline{\tau}/WW^{*} /ZZ^{*}$ processes are estimated to be approximately $0.55\%$, $8.59\%$, $3.03\%$, $2.85\%$, $1.58\%$ and $15.81\%$, respectively. Compared to a previous analysis \cite{Bai_2020}, which reported statistical uncertainties of $1.1\%$, $10.5\%$ and $5.4\%$ for the branching fractions of $H\to b\overline{b} /c\overline{c} /gg$ process, the PFN method achieves higher precision in a single execution, due to its better performance and deeper data exploitation. By increasing the transverse momentum resolution by 2\% to account for differences between real data and simulated samples, the systematic uncertainties for the branching fractions are estimated to be 0.21\%, 3.88\%, 2.74\%, 1.39\%, 0.18\% and 19.09\% for $ b\overline{b} /c\overline{c} /gg/\tau\overline{\tau}/WW^{*} /ZZ^{*} $ final states, respectively. This study achieves highly precise measurements of decay branching fractions of Higgs, helping to increase the understanding of the properties of the Higgs boson and further testing of the Standard Model.

\section*{Acknowledge}
\textit{We would like to The authors present special thanks to CEPC higgs physics working group for useful discussion and advice. The authors thank the IHEP Computing Center for its firm support. This work is supported in part by National Key R\&D Program of China under Contracts No. 2022YFE0116900 and the Basic Science Center Program No. 12188102 by National Natural Science Foundation of China.}



\bibliographystyle{pisikabst}
\bibliography{bibfile}

\end{document}